\documentclass[fleqn,usenatbib]{mnras}

\usepackage{newtxtext,newtxmath}

\usepackage[T1]{fontenc}

\DeclareRobustCommand{\VAN}[3]{#2}
\let\VANthebibliography\thebibliography
\def\thebibliography{\DeclareRobustCommand{\VAN}[3]{##3}\VANthebibliography}

\usepackage{graphicx}	

\title[The Rise and Fall of Dust]{The Rise and Fall of Dust in the Universe}

\author[S.A. Eales et al.]{Stephen Eales,$^{1}$\thanks{E-mail: steve.a.eales@gmail.com}
and Bradley Ward$^{1}$
\\
$^{1}$Cardiff Hub for Astrophysics Research and Technology, School of Physics and Astronomy, Queen's Buildings,\\
5 The Parade, Cardiff CF24, 3AA, UK
}

\date{ }

\pubyear{2015}

\begin{document}
\label{firstpage}
\pagerange{\pageref{firstpage}--\pageref{lastpage}}
\maketitle

\begin{abstract}

We estimate how the mean density of dust in the universe varies with redshift, using
submillimetre continuum observations and a method designed to minimise the effect of dust temperature. We have used the Herschel-ATLAS
to show that the median temperature of dust in galaxies is $\rm \simeq 22\ K$ and does not vary significantly with
redshift out to $z=1$. With this as our estimate of the mass-weighted dust temperature, we have used an 850-$\mu$m survey of the COSMOS field
to estimate the mean density of dust in 10 redshift bins over the range $0 < z < 5.5$.
We find that the mean density of dust increased by a factor of $\simeq$10 from $z=5$ to $z=2$, declined
slightly to $z=1$, and then steeply to the present day. The relationship between the mean density of dust and redshift is similar to the relationship
between the mean star-formation rate and redshift, although the increase for the former is steeper from $z=5$ to $z=2$.
We have also used the submillimetre measurements to estimate the mean density of
gas over the same redshift range. The values we estimate for the dust-traced gas are much lower and with a different redshift dependence
than those for estimates of the mean density of atomic gas but similar to those for estimates of the mean density of the CO-traced gas.
We find that the depletion time for the dust-traced gas in the universe as a whole declines with redshift in the same way
seen for individual galaxies.
\end{abstract}

\begin{keywords}
galaxies: evolution -- galaxies: ISM -- submillimetre: galaxies
\end{keywords}



\section{Introduction}

Dust is one of the most importance substances in the universe, being the critical coolant in star-formation regions and
being responsible for the formation of molecular
hydrogen, which in turn makes it indirectly responsible
for the creation of stars and planets. Apart from these important astrophysical roles and its intellectual interest, dust is also a valuable astronomical
tool as a tracer of the ISM. CO, the traditional tracer of the 
molecular phase for the last 50 years, has many well-known
problems, and the use of the continuum emission from dust as an alternative method of tracing the molecular phase
of the ISM \citep{Eales_2012,Scoville_2014} has many advantages:
dust grains are more robust than CO molecules; it is much easier to detect the dust in
galaxies than the CO ($\sim 10^6$ detections of galaxies in the {\it Herschel} archive); 
and, if the metal abundance in the galaxy is known, the relationship
between the dust-to-gas ratio and metal abundance is much simpler than the relationship between the CO X-factor and
metal abundance \citep{Bolatto2013,remy-ruyer2014}.
Observations of the dust in large samples of galaxies are already giving valuable insights into
galaxy evolution. For example, the rapid low-redshift evolution seen in the submillimetre luminosity function
\citep{dye2010} and the dust-mass function \citep{dunne2011} is not reproduced by cosmological simulations 
\citep{Eales_2018,Millard2021}.

In this paper, we do not consider the dust content of individual galaxies, but we instead describe an
attempt to determine how the mean density of dust depends on redshift. There are three main
methods that have been used to derive this relationship: i) from
the absorption-line systems in quasars, for which the line ratios can be used to estimate both the
metal abundance and the fraction of metals locked up in the dust; ii) from the dust reddening of the quasars with absorption-line
systems; iii) from the continuum emission of the dust in galaxies \citep{peroux2020}. The main disadvantage of the first two methods is that the dust
in the absorption-line system reduces the optical brightness of the quasar, creating a selection bias
against quasars behind dust-rich absorption-line systems being included in quasar catalogues \citep{peroux2020}.

In this paper, we apply the third of the methods. This does not suffer from the potential selection
bias of the other two but has problems of its own. The method relies on the assumption that the
constant of proportionality between the dust mass and the submillimetre luminosity is independent of redshift.
We cannot be completely confident of
this assumption, although there is evidence the dust emissivity index, which is dependent on the structures
and chemistry of the dust grains, does not change with redshift (\citet{Ismail_2023,Witstok_2023}, Ward et al. MN, submitted). This, however, is only a potential problem. An actual problem is that the emission from the dust also depends
on the temperature of the dust.

Something that limits the accuracy of all attempts to measure the temperature of the dust in galaxies is
that warm dust emits more radiation than the same mass of cold dust. This means that the
temperature estimated from the best-fitting SED, the luminosity-weighted mean dust temperature, 
will always be higher than the more meaningful mass-weighted mean dust temperature unless the temperature
of the dust is everywhere the same in the galaxy \citep{eales1989}. 
The effect of this bias on the estimate of the dust mass depends
on the rest-frame wavelength of the flux density used to estimate the dust mass.
The effect is less for wavelengths on the long-wavelength (Rayleigh-Jeans) side of the peak
in the dust SED, since at infinite wavelength the continuum emission depends on only the
first power of dust temperature. The effect
of the temperature bias can therefore be reduced by estimating the
dust masses from a survey carried out as long a wavelength as
possible. This is the approach we have followed in this paper,
in which the dust masses have been estimated from a survey
at 850 $\mu$m.

This paper is arranged as follows. In Section 2 we describe the method. Section 3 gives the results.
We discuss the results in Section 4 and list our conclusions in Section 5. We use
the cosmological parameters given in \cite{Planck_2015}.

\section{Methods}

\begin{figure}
	\includegraphics[width=\columnwidth]{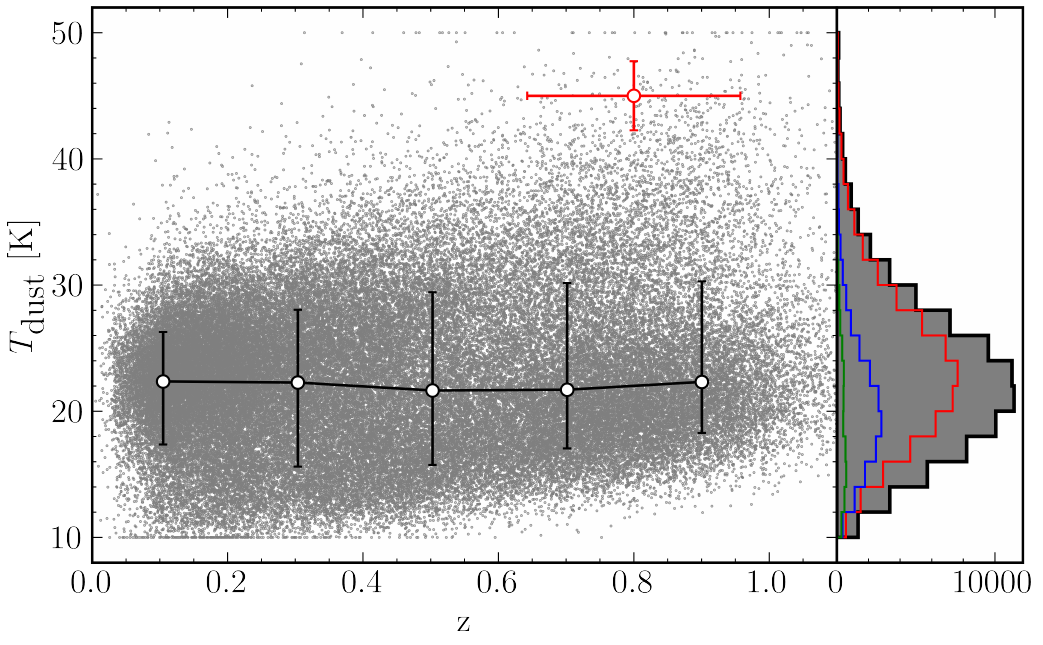}
	\caption{Plot of dust temperature against redshift for 81,895 galaxies in the SGP
 field of the Herschel ATLAS. The red cross shows the typical $\pm1\sigma$ uncertainty
 in the photometric redshift and
 temperature estimate. The open circles show the median
 dust temperatures in five redshift bins, each of width 0.2, with the error bar showing
 16th to 84th percentiles of the temperature distribution. The panel to the right shows the
 histogram of dust temperature for all galaxies (filled-in histogram), galaxies with
 $S/N>4$ in one band (red), in two bands (blue) and in
 three bands (green).}
	\label{fig:dust temperatures}
\end{figure}

\subsection{The Temperature of the Dust}

The SED of the dust in a typical galaxy has a peak in the rest-frame
at $\rm \lambda_{peak} \simeq 100\ \mu m$. If a small amount of hot dust is
added to the galaxy, it will have most effect on the
SED at wavelengths $\rm < \lambda_{peak}$.
One might therefore hope to get a more accurate
estimate of the mass-weighted dust temperature from fitting a modified
blackbody to the galaxy's flux measurements if the fit only
uses flux measurements at $\rm \lambda > \lambda_{peak}$.
Empirically, this seems to be true. Estimates of dust temperature for
galaxy samples in which flux measurements at shorter wavelengths
are included (e.g. \citet{Dunne_2000,Vlahakis2005} are much higher than ones in
which only flux measurements at longer 
wavelengths are included (e.g. \citet{Eales_2012,Cortese2014}.
There is also evidence for individual galaxies that inclusion of
a flux measurement at a wavelength shorter than the peak
increases the estimate of the dust 
temperature \citep{Smith2010,Smith2012}.

With this in mind, we have estimated dust temperatures from the
galaxies
discovered in the Herschel ATLAS (H-ATLAS),
a survey of 660 deg$^2$ carried out with Herschel Space Observatory 
\citep{Eales_2010,valiante2016,bourne2016,smith2017,maddox2018,furlanetto2018,ward2022}
at 100, 160, 250, 350 and 500 $\mu$m. 
The advantages of this survey are that it provides long-wavelength photometry
for a very large number of galaxies. The H-ATLAS photometry is most sensitive
at the three longest wavelengths and in the rest-frame these wavelengths
are at $\rm > \lambda_{peak}$ at the highest redshifts
we consider.

We used the catalogue of sources from the H-ATLAS field at the south galactic pole, which covers 317.6 deg$^2$ 
\citep{maddox2018,ward2022}. We used the subset of 81,895 sources for which
there is both a probable counterpart and a redshift \citep{ward2022}. 
The redshifts are photometric redshifts estimated from the available
multi-wavelength images for this field \citep{ward2022}.
Each of the sources in this sample was detected at $>4\sigma$ at at least one of the following wavelengths:
250$\mu$m, 350$\mu$m, 500$\mu$m. Each source has a flux measurement at all five
H-ATLAS wavelengths.

We fitted the standard modified blackbody equation to the five flux densities available
for each source:

\begin{equation}
    F_{\nu} \propto B(\nu) \nu^{\beta}
	\label{eq:mbb}
\end{equation}

\noindent in which $\beta$ is the dust emissivity index. We used a value of
$\beta$ of 2, which is typical of the values found in most recent studies of
high-redshift galaxies (\citet{Bendo_2023,Ismail_2023,Witstok_2023}; Ward et al., MN, submitted).

Figure \ref{fig:dust temperatures} shows the median dust temperature of the sources in five redshift
bins. Table \ref{tab:temperatures} lists the median values 
and the 16th to 84th percentiles of the temperature distribution in each bin.
Over the redshift range $0<z<1$ there is no evidence of any evolution
in dust temperature. The median dust temperature in all bins is $\simeq$22 K. The right-hand panel shows the histograms of dust temperature
for sources for which there are detections at $>4\sigma$ at one, two or three of
the {\it Herschel} wavelengths. The panel implies that the median dust temperature
would be lower if we only considered sources with at least two or three significant detections.
However, this 
might be a selection effect because it is more likely there will be an additional
significant detection at $\lambda > 250 \mu$m if the galaxy is colder. We therefore decided to use the median
dust temperature derived from the entire sample.

\begin{table}
	\centering
	\caption{The Median Dust Temperature}
	\label{tab:temperatures}
	\begin{tabular}{lc}
		\hline
		Redshift  & $T_d$ \\
          & (K) \\
		\hline
        0.0<z<0.2 & $22.4_{-5.0}^{+3.9}$ \\
        0.2<z<0.4 & $22.3_{-6.7}^{+5.7}$ \\
        0.4<z<0.6 & $21.7_{-5.9}^{+7.7}$ \\
        0.6<z<0.8 & $21.7_{-4.7}^{+8.5}$ \\
        0.8<z<1.0 & $22.3_{-4.1}^{+8.0}$ \\
		\hline
	\end{tabular}
\end{table}

There is no consensus in the literature about whether 
dust temperature increases with redshift at $z>1$, with some studies finding
positive evolution \citep{Drew_2022,Magnelli_2014,Bethermin_2015,Zavala_2018,Schreiber_2018} and others
no evolution \citep{Barger_2022,Dudzeviciute_2020,Lim_2020}.
A fundamental limit to these studies is that they all measure the luminosity-weighted
rather than the mass-weighted
dust temperature. We have 
recently estimated the dust temperature in a sample of 100 dusty star-forming galaxies
in the redshift range $2 < z < 6$ (Ward et al. MN, submitted), in which we have used only photometry on the long-wavelength side of
the dust peak {\bf (see above)}. We found no evolution in the temperature of the dust over this redshift range

Models that use simulations of galaxies over the redshift range $2 < z < 6$ from the
{\it Feedback in Realistic Environments} (FIRE) project show (a) that the luminosity-weighted
dust temperature ($\rm T_{eff}$ in their nomenclature) is often $\simeq$10 K higher than the mass-weighted dust temperature
and (b) that the mass-weighted dust temperature
does not evolve strongly with redshift \citep{Liang_2019}. 68\% of the
simulated galaxies in this study have mass-weighted dust temperatures in the range $\rm T_d = 25\pm5 K$, which is not too far off
the values we measure at $z<1$.

Given the results of the FIRE simulations and the lack of any clear contrary observational evidence,
in the rest of this paper we have assumed that the median dust temperature at all redshifts is 22 K, although
we have explored the consequences if this assumption is not correct.

\subsection{Calculating the Dust Density}

In our analysis, we have combined two sets of results for the COSMOS field.
First, \cite{davidzon2017} have measured the stellar mass function (SMF) in the field
in 10 redshift slices over the redshift range $0 < z < 5.5$. Second, \cite{Millard_2020} used a stacking analysis on the S2COSMOS 850-$\mu$m survey of the
COSMOS field \citep{Simpson_2017} to measure
the mean 850-$\mu$m flux of 64,684 galaxies, separated into bins
of stellar mass and redshift, in the field. We have
combined these two datasets to estimate the mean density of dust in each redshift bin.

We have calculated the dust mass from the flux density
at 850 $\mu m$ \citep{Millard_2020} using the equation:

\begin{equation}
    M_d = \frac{<S_{850\mu m}> D_L^2}{(1+z) B(\nu_e) \kappa(\nu_e)} 
	\label{eq:md}
\end{equation}

\noindent in which $\nu_e$ is the frequency in the rest frame: $\nu_e = (1+z) \nu_{850\mu m}$. $D_L$ is luminosity distance, $<S_{850 \mu m}>$ is the mean flux density
given by \cite{Millard_2020}, and the
dust mass-opacity coefficient, $\kappa(\nu)$, is given by:

\begin{equation}
    \kappa(\nu) = \left(\frac{\nu}{\nu_{850 \mu m}} \right)^{\beta} \kappa_{850 \mu m} 
	\label{eq:kappa}
\end{equation}

\noindent We assumed a value for $\beta$ of 2 (see above), a value for the dust temperature of
22 K, and the widely used value for the mass-opacity coefficient of
$\kappa_{850\mu m} = 0.077\ m^2\ kg^{-1}$ \citep{James_2003,dacunha2008,dunne2011}.
Before calculating the dust mass, we corrected the flux density and the
dust temperature for the effect of the cosmic microwave background using the method in \citet{daCunha_2013} (equations 12 and 18 in that paper).

\begin{figure}
	\includegraphics[width=\columnwidth]{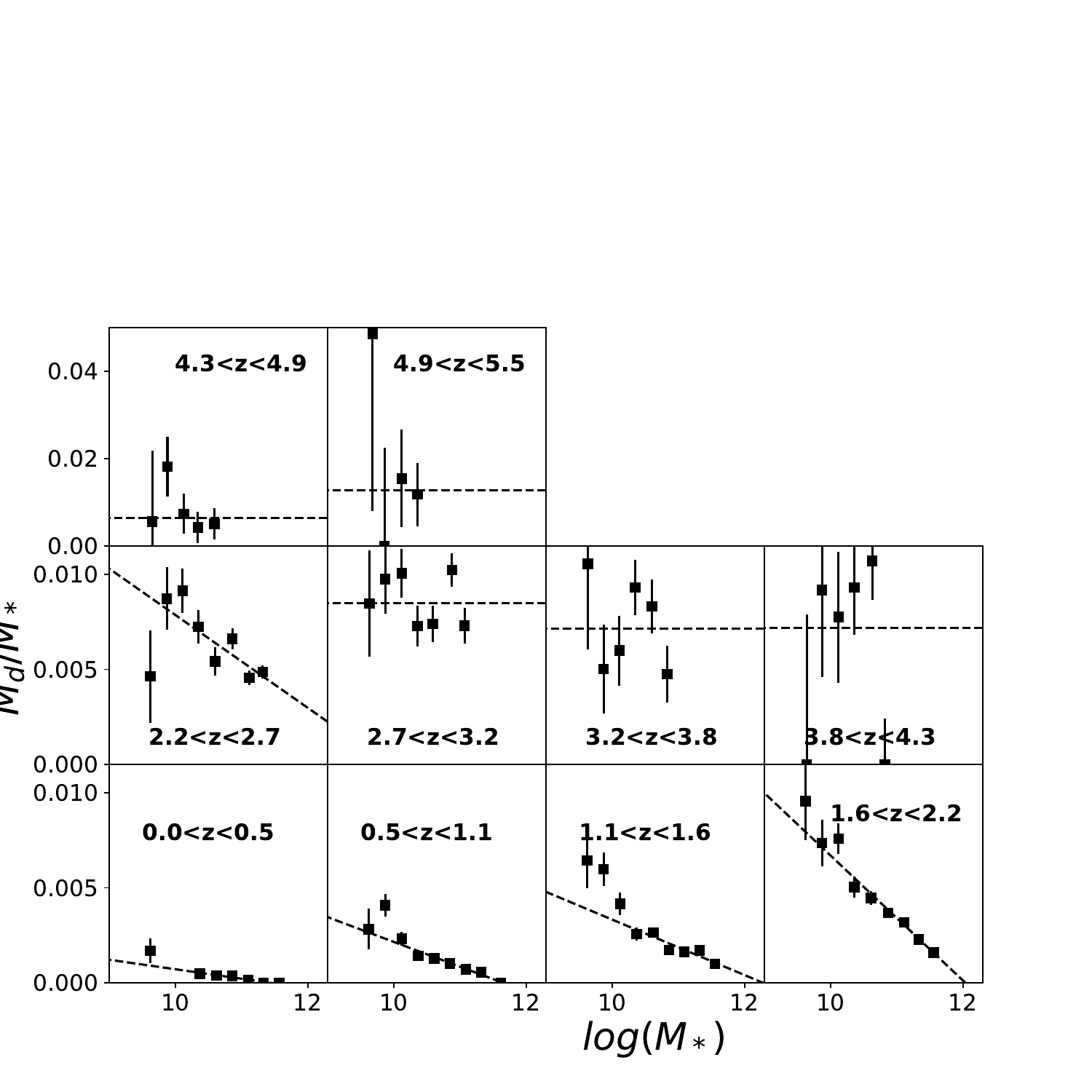}
	\caption{Plot of the ratio of dust mass to stellar mass against stellar mass
 for the redshift ranges used by \citet{Millard_2020}. 
 For the five low-redshift bins, the dashed lines show the best-fitting
 linear relationships (equation 4; Table 2). For the five high-redshift bins
 the dashed lines show the weighted mean value of the mass ratio (Table 2).
 }
	\label{fig:dust masses}
\end{figure}

Figure \ref{fig:dust masses} shows the ratio of the mean value of the dust mass to the median stellar mass for each redshift bin plotted against stellar mass,
using the mean redshift and median stellar mass for each bin
(see Table D1 of \citet{Millard_2020}.
For each of the bins, we fitted the
following relationship to the data points:

\begin{equation}
    \frac{M_d}{M_s} = a log_{10}M_s + b
	\label{eq:fit 1}
\end{equation}

\noindent Table \ref{tab:dust mass stellar mass ratio} lists our estimates of $a$ and $b$ and the values
of the weighted mean ratio of dust mass to stellar mass for each of the redshift bins. There is
a clear relationship between the mass ratio and stellar mass for the five low-redshift bins, with
the best-fitting relationship shown in the figure. For the other five bins, the dashed line in the
figure shows the weighted mean value of the mass ratio.

We calculated the mean density of dust, $\rho_i$, in the i'th bin of stellar mass and redshift from
the equation:

\begin{equation}
    \rho_i = \frac{ \int_{z_L}^{z_U} \int_{M_L}^{M_U} \phi(M_s) M_d(z) dM_s dV}{\int_{z_L}^{z_U} dV}
	\label{eq:dust density}
\end{equation}

\noindent in which $\rm M_s$ is the mass of stars in a galaxy and the dust mass is calculated
from the mean flux density given in \cite{Millard_2020} using equation 2. We did not interpolate
across a bin using the relationships given in Table 2 because this is not consistent with
the stacking method used in the earlier paper.

The mean density of dust in each redshift bin is
then given by:

\begin{equation}
    \rho = \sum_i \rho_i + c
	\label{eq:dust density 2}
\end{equation}

\noindent in which the sum is over all the stellar-mass bins for that redshift slice and $c$ is a correction factor
to include the contribution to the dust density from galaxies outside
the range covered by the stacking analysis \citep{Millard_2020}. We chose to estimate the mean density of dust for galaxies with a stellar mass in the range $9.0 < log_{10}(M_*) < 11.75$
and calculated $c$ using the stellar mass function and the relationships shown in Figure 2.
For the five low-redshift bins, we used the best-fitting
linear relationship. For the five high-redshift bins, for which there is not a clear trend
between the mass ration and stellar mass, we used the weighted mean value of the mass ratio to calculate $c$, but
we also made an estimate of $c$ using the best-fitting linear relationship. We used the
difference between the two estimates of $c$ as a systematic error, which we added in quadrature to
the regular error. The contribution of $c$ (Table 3) to the mean dust density only exceeds 20\% at $z>3$. 
We calculated the statistical errors in the mean dust density by combining in quadrature the errors in the mean dust density in each stellar mass/redshift bin which arise
from the error in the mean 850-$\mu$m flux density for that bin.

\begin{table}
	\centering
	\caption{Fits to $M_d/M_s$ versus $M_s$ (Fig. \ref{fig:dust masses})}
	\label{tab:dust mass stellar mass ratio}
	\begin{tabular}{lccc} 
		\hline
		Redshift  & a & b & $<M_d/M_*>$ \\
		\hline
0.0<z<0.5 & -0.00050 & 0.0057 & 0.00014$\pm$0.00002  \\
0.5<z<1.1 & -0.00130 & 0.0152 & 0.00076$\pm$0.00004 \\
1.1<z<1.6 & -0.00146 & 0.0179 & 0.0018$\pm$0.00006 \\
1.6<z<2.2 & -0.00329 & 0.0396 & 0.0031$\pm$0.0001 \\
2.2<z<2.7 & -0.00244 & 0.0323 & 0.0054$\pm$0.0002 \\
2.7<z<3.2 & -0.00058 & 0.0146 & 0.0085$\pm$0.0004 \\
3.2<z<3.8 & -0.00105 & 0.0181 & 0.0071$\pm$0.0007 \\
3.8<z<4.3 & -0.00459 & 0.0553 & 0.0072$\pm$0.0012 \\
4.3<z<4.9 & -0.01115 & 0.1217 & 0.0065$\pm$0.0021 \\
4.9<z<5.5 & -0.00866 & 0.1015 & 0.0128$\pm$0.0058 \\
		\hline
	\end{tabular}
\end{table}

\begin{figure*}
	\includegraphics[width=\textwidth]{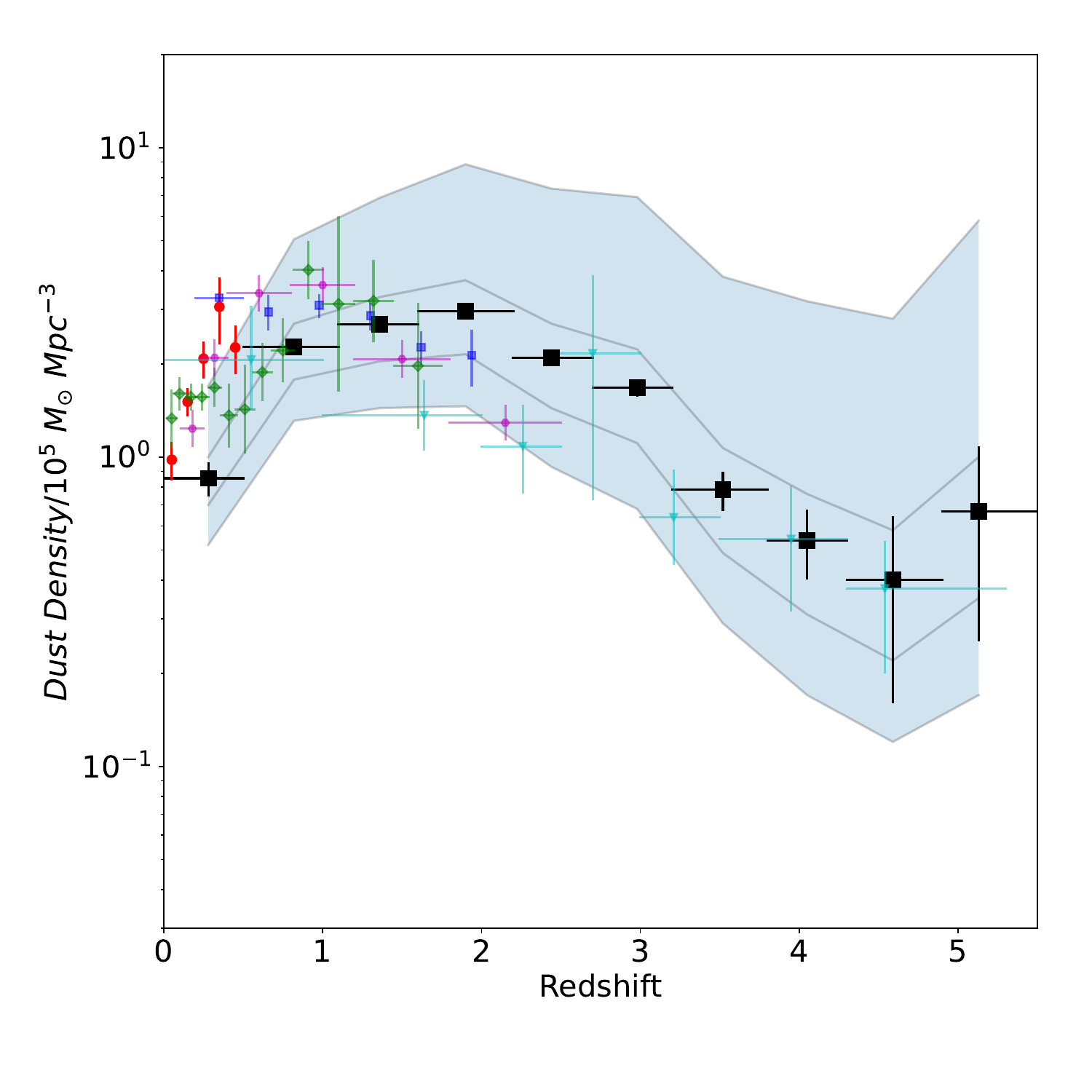}
	\caption{Mean density of dust in the universe versus redshift. 
 The large black squares are our estimates (Table 3). The four grey lines link 
 our estimates if we repeat the analysis using dust temperatures, from bottom to top, 
 of 30 K, 25 K, 20 K and 15 K. The other symbols show previous estimates of the dust density: red \citep{dunne2011}, blue \citep{menard2010,menard2012}, green \citep{driver2018}, 
 mauve \citep{pozzi2020} and cyan \citep{peroux2020}. We 
 have taken the values for all except the first from the useful compilation of \citet{peroux2020}.}
	\label{fig:dust density}
\end{figure*}

\section{Results}

Figure \ref{fig:dust density} shows our estimates of the mean density of dust in each redshift bin,
with values listed in Table \ref{tab:dust density}. 
To assess the possible systematic errors caused by errors in our assumption about dust temperature, we also
estimated the mean density of dust using dust temperatures
of 30 K, 25 K, 20 K and 15 K. These estimates are linked by the grey lines in the figure. 
The
grey lines show that the main consequence if our assumption about the mean temperature of dust is incorrect
is a change in the overall amplitude of the relationship rather than its shape,
which is a result of the long wavelength used in the stacking analysis (on the Rayleigh-Jeans tail of a modified blackbody, emission is
proportional to dust temperature). These lines also make it clear what would happen if our assumption that the
mass-weighted temperature of dust is independent of redshift is incorrect.
If, as some have claimed (\S2.2), dust temperature increases with redshift, the decline in the mean density of dust
at $z>2$ would be steeper than
we have estimated.

Our analysis shows that the mean density of dust in the universe increased by a factor of $\simeq$10 from $z=5$ to $z=2$. 
There is then marginal evidence (1.6$\sigma$) for a decline from the bin for $1.6<z<2.2$ to the bin for
$1.1<z<1.6$. The mean dust density then declined steeply to the present day. We have also plotted in the figure some previous estimates of the 
mean dust density. The previous observations, considered as a whole, agree with
our results, but there are interesting discrepancies between the different datasets.

The estimates that are most comparable to ours are those of \cite{dunne2011} and \cite{pozzi2020}, who also used
the submillimetre emission from dust to estimate the mean density of dust. The three datasets agree well about the rapid increase at $z < 1$, but our estimates are systematically lower than the
other two. The obvious explanation of the difference in the one case is that \cite{pozzi2020} used
a value for the mass-opacity coefficient a factor of $\simeq$2 lower than we did, which will increase their
estimates relative to ours by the same factor. Even with this correction, their values begin to diverge substantially
from ours at $z>1$, but at high redshifts their estimates are likely to be less reliable because 
their short selection wavelength (160 $\mu$m) means that their mass estimates will be highly dependent on dust temperature.
The reason for the difference in the other case is less clear because if $\beta=2$, we and \cite{dunne2011} used the
same values for the mass-opacity coefficient. However, because we used samples selected at different wavelengths
(850 and 250 $\mu$m), a small difference in $\beta$ would lead to a large change in the mass-opacity coefficient.
If the mass-opacity coefficient is fixed at
850 $\mu$m, an increase in $\beta$ of 0.3, for example, would increase the mass-opacity coefficient at 250 $\mu$m by
$\simeq$44\%, which would decrease the mass estimates
by the same factor. 

The approach used by \cite{driver2018} is a hybrid one in which all of their
galaxies have submillimetre observations but most do not have signficant submillimetre detections, with the dust masses being estimated
from fits to the multi-wavelength photometry (UV to submm) with the MAGPHYS modelling programme \citep{dacunha2008}. They do not
find the rapid low-redshift evolution in the mean density of dust seen by us and in the other two submillimetre papers.

Our estimates agree well at $z >2.5$ and $z < 1$ with the estimates by \cite{peroux2020} from the line ratios in quasar absorption-line systems,
but they are much higher than the absorption-line estimates in the intervening redshift range, which is the redshift range in which
we find the mean dust density is at its peak. We speculate that the difference
might be caused by the selection effect from dust in the absorption line systems (\S 1) being
greatest in this redshift range.
Our estimates do not agree well with the estimates from the reddening of quasar
absorption-line systems \citep{menard2010,menard2012} at any redshift.

\section{Discussion}

\begin{figure*}
	\includegraphics[width=\textwidth]{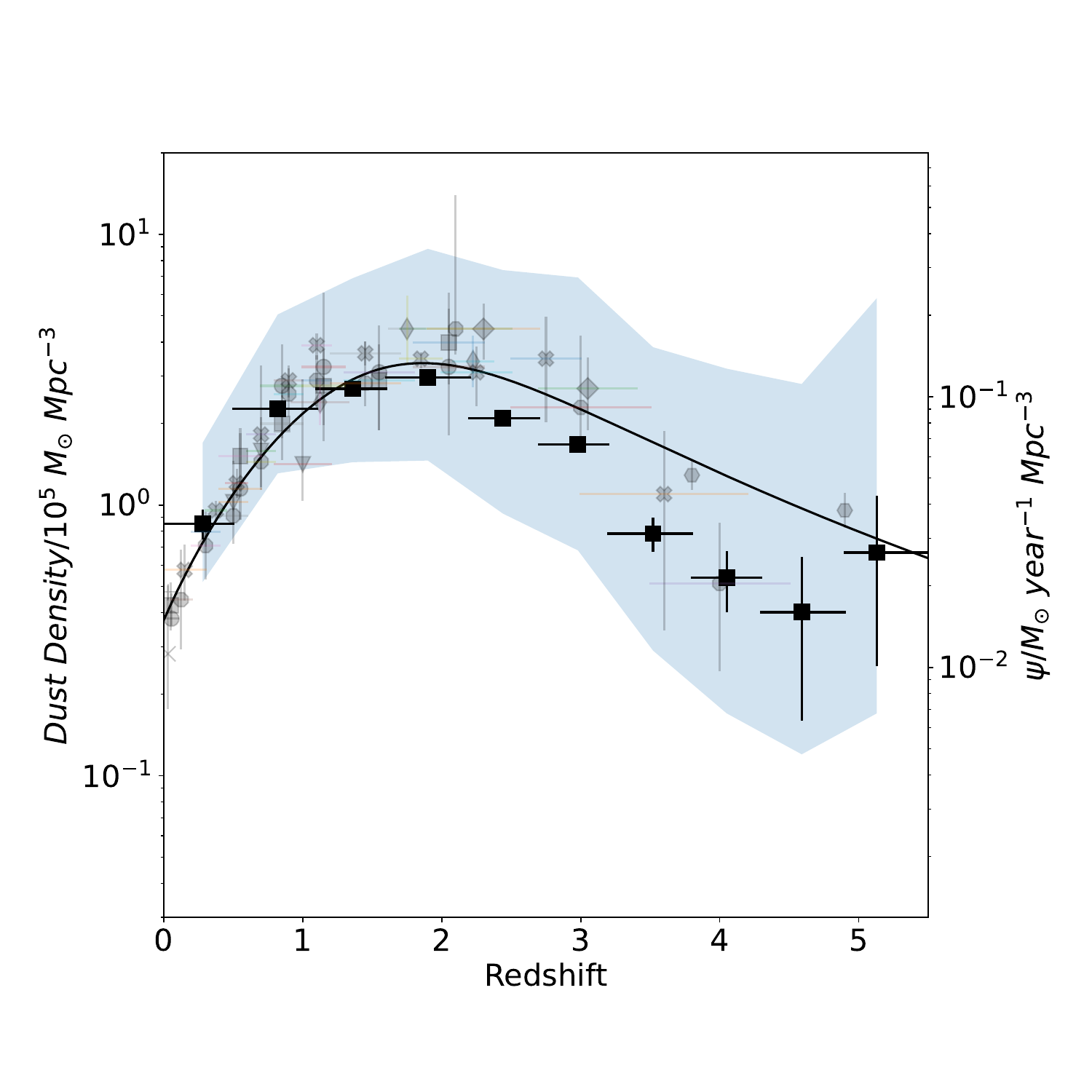}
	\caption{Mean density of dust versus redshift with overplotted estimates
 of the mean star-formation rate in the universe. The large black squares are our estimates of the mean density of
 dust (Table \ref{tab:dust density}). The other symbols are estimates of the mean star-formation rate in the universe \citep{madau2006}, with the solid line
 showing a relationship that is a good fit to these measurements \citep{madau2006}.
 The vertical position of the star-formation measurements has been chosen to roughly coincide with the
 dust measurements.}
	\label{fig:dust density with sfr}
\end{figure*}

Figure \ref{fig:dust density with sfr} shows our measurements of the mean density of dust plotted against redshift with overplotted a compilation of measurements
of the mean star-formation rate in the universe \citep{madau2006}. The line is a relationship that is a good fit to the
measurements of the mean star-formation rate \citep{madau2006}. We have chosen the offset between the mean star-formation rate
(shown on the right) and the mean dust density (shown on the left) so that the two sets of measurements roughy coincide.
Although the vertical positioning is arbitrary, the figure shows that the redshift dependences of the two relationships
are similar. Both the mean density of dust and the mean star-formation rate peak at $z\sim2$, and the two depend
on redshift in similar ways except that the decline with redshift at $z>2$ is steeper for the dust than for the star-formation rate.

\begin{figure}
	\includegraphics[width=\columnwidth]{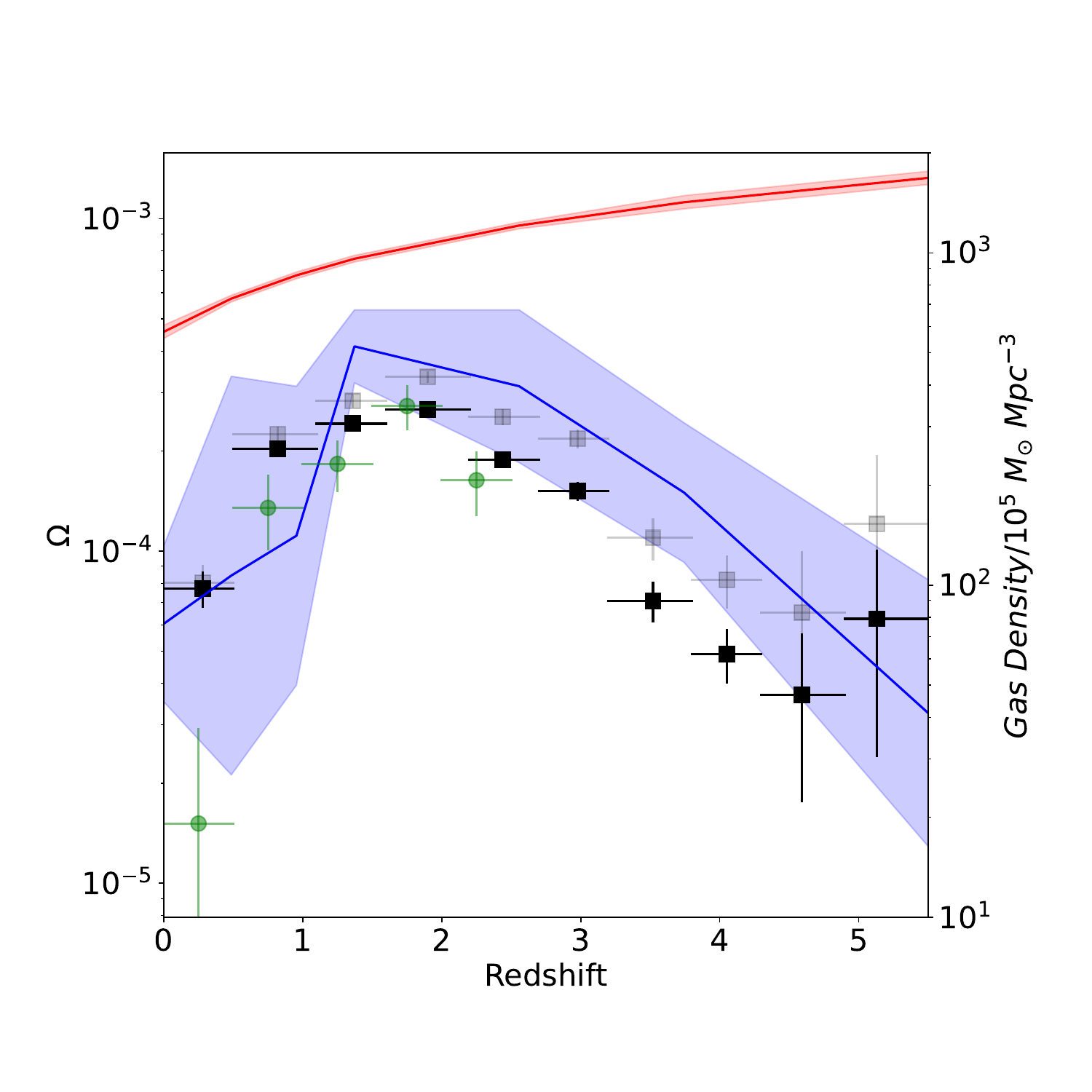}
	\caption{Mean density of hydrogen plotted against redshift, expressed as a fraction of the universe's critical density
 on the left-hand axis and as a direct density on the right-hand axis. The black squares are our estimates using
 the relationship between hydrogen mass and submillimetre luminosity (equation 7). The coloured lines and bands show the best current
 estimates, and their uncertainties, of the mean density of atomic hydrogen (red) and of the molecular hydrogen traced by CO
 (blue) \citep{peroux2020}.
 The grey squares are our estimates if
 we repeat our analysis with a dust temperature of 18 K rather than 22 K. The green symbols show estimates from
 another study that used the continuum emission from the dust to trace the gas \citep{Garratt2021}}
	\label{fig:gas density}
 \end{figure}

We have also used our results to estimate how the mean density of gas in the universe varies with redshift (\S 1). To estimate this, it is not
necessary to know the values of the dust mass-opacity coefficient or the gas-to-dust ratio. Instead, in a similar way to the other ISM 
tracers, CO molecules and carbon atoms \citep{Dunne_2022}, the mass of the molecular
ISM can be estimated directly
from the submillimetre luminosity:

\begin{equation}
    M_{H_2} = \alpha_{submm} L_{submm}
	\label{eq:cal2}
\end{equation}

\noindent as long as the constant of proportionality in the equation is known. There are several estimates of $\alpha_{submm}$ in the
literature \citep{Scoville_2014,tacconi2020,Dunne_2022}, all of which are fairly consistent. We have used the
value in \cite{Dunne_2022} because these authors used a method that produced mutually consistent estimates of the values of $\rm \alpha$ for the three standard
tracers: CO, CI and
dust. We used a value of $\rm \alpha_{850 \mu m} = 1.066 \times 10^{-13}\ M_{\odot}\ Hz\ W^{-1}$, a factor 1.36 lower than
the value given in \cite{Dunne_2022}, in order that our mass estimates would not include the contribution of helium.
Figure \ref{fig:gas density} shows our results. The green symbols show the estimates from another recent study
that also used the continuum emission from the dust to trace the gas \citep{Garratt2021}. This study found a similar
rise and fall of the gas density with redshift that we see, although there are significant differences between our and
their estimates at some redshifts.

The coloured lines and bands in the figure show the best current
 estimates, and their uncertainties, of the mean density of atomic hydrogen (red) and of molecular hydrogen
 (blue) \citep{peroux2020}. The molecular hydrogen estimates come from observations of another
 tracer, CO, and were made with a value for $\alpha_{CO}$ of 
 3.6 $\rm M_{\odot}\ (K\ km\ s^{-1}\ pc^2)^{-1}$ \citep{walter2014,decarli2016,decarli2019,riechers2019}. Although
 this is similar to the value found by \cite{Dunne_2022}, 4.0 $\rm M_{\odot}\ (K\ km\ s^{-1}\ pc^2)^{-1}$, the latter value
 includes helium while the other value does not. 
 Since \cite{Dunne_2022} obtained estimates of the values of $\rm \alpha$ for the three tracers in a self-consistent way, we have scaled the CO measurements to a value for $\alpha_{CO}$ of 2.94 $\rm M_{\odot}\ (K\ km\ s^{-1}\ pc^2)^{-1}$,
 which is the value given by \citet{Dunne_2022} with the contribution of the helium removed.
  
 There are large differences, both in redshift dependence and normalisation, between our estimates  and the relationship
 for atomic hydrogen. The similarity in the
 shapes between the relationships derived from the two tracers, CO and dust, imply that cosmic dust is mostly found in molecular gas.
 
 The dust-traced gas points agree well with the CO-traced gas estimates at $z < 1$ but are significantly lower at higher
 redshift. One possible explanation is that our estimate of the mass-weighted dust temperature is
 still too high. The faint points in the figure show the effect of using a dust temperature of 18 K rather
 than 22 K, which is enough to make the estimates from the CO and dust consistent within the errors. Another possible
 explanation is that the ratio $\rm \alpha_{CO}/\alpha_{submm}$ depends on redshift, although \cite{Dunne_2022} concluded
 that this ratio is roughly constant over the redshift range $0 < z < 6$.

The depletion time is the time needed for star formation to consume all the gas in a galaxy if no more is accreted. It is given by:

\begin{equation}
    \tau_{dep} = \frac{M_{gas}}{SFR}
	\label{eq:cal}
\end{equation}

\begin{figure}
	\includegraphics[width=\columnwidth]{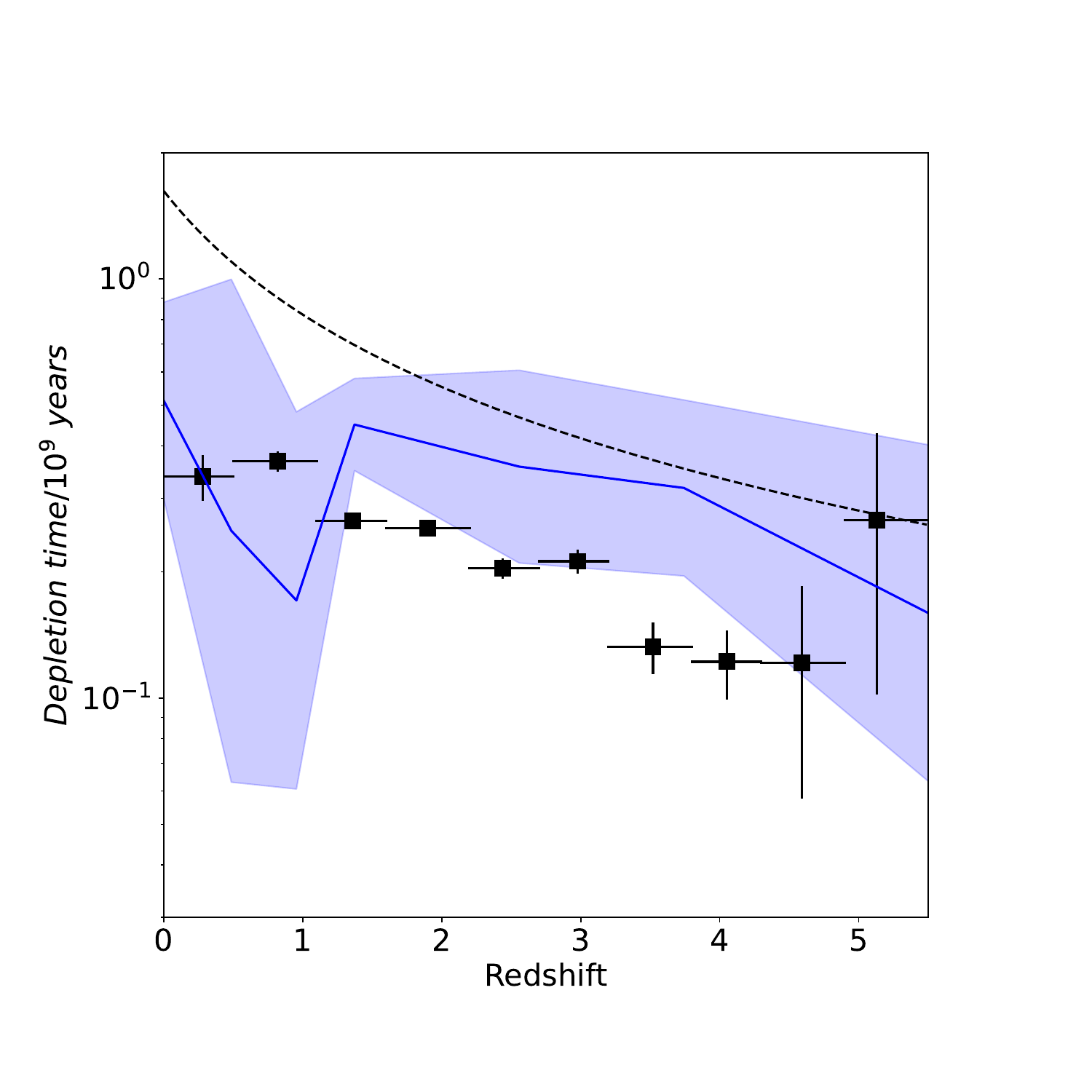}
	\caption{Depletion time versus redshift.
 The black squares show the estimates for the dust-traced gas. The blue line and band show the
 estimates and their uncertainty for the CO-traced gas \citep{peroux2020}. The dashed
 line shows an estimate for individual galaxies on the star-forming main
 sequence and with $log_{10}(M_*) = 10.5$} \citep{tacconi2020}.
 
	\label{fig:depletion times}
 \end{figure}

\noindent It is possible to use the estimates of the mean density of gas in Figure \ref{fig:gas density} and the estimates
of the mean star-formation rate to calculate the depletion time for the gas in the universe as a whole. We have used equation (15) in \cite{madau2006}
to estimate the mean star-formation rate at each redshift. Figure \ref{fig:depletion times} shows our estimates of the depletion times
for both the CO-traced gas and the dust-traced gas. The dashed line shows the relationship between depletion time
and redshift derived for galaxies on the star-forming main sequence and with
$log_{10}(M_*) = 10.5$, which makes it comparable to the galaxies
in our sample \citep{tacconi2020}. The relationship for
the dust-traced gas and that derived from observations of individual galaxies show very similar declines with redshift, although there is an offset
as there was for
Figure \ref{fig:gas density}. As with the previous figure, much of the
discrepancy might be explained if the temperature of the dust is colder than we have assumed.

\begin{table}
	\centering
	\caption{The Mean Density of Dust}
	\label{tab:dust density}
	\begin{tabular}{lccc} 
		\hline
		Redshift  & $<z>$ & $<\rho_{dust}>$ & correction$^a$\\
   & & $(\rm 10^5\ M_{\odot}\ Mpc^{-3})$ & $\rm (10^5\ M_{\odot}\ Mpc^{-3})$ \\
		\hline
0.0<z<0.5 & 0.28 & 0.85$\pm$0.11 & 0.10 \\
0.5<z<1.1 & 0.82 & 2.27$\pm$0.13 & 0.22 \\
1.1<z<1.6 & 1.36 & 2.69$\pm$0.12 & 0.20 \\
1.6<z<2.2 & 1.90 & 2.96$\pm$0.12 & 0.34 \\
2.2<z<2.7 & 2.44 & 2.09$\pm$0.12 & 0.32 \\
2.7<z<3.2 & 2.98&  1.68$\pm$0.11 & 0.30 \\
3.2<z<3.8 & 3.52 & 0.78$\pm$0.11 & 0.22 \\
3.8<z<4.3 & 4.05 & 0.54$\pm$0.14 & 0.19 \\
4.3<z<4.9 & 4.59 & 0.40$\pm$0.24 & 0.15 \\
4.9<z<5.5 & 5.13 & 0.67$\pm$0.41 & 0.26 \\
		\hline
	\end{tabular}
 \begin{flushleft}
$^a$The correction has been included in the estimate in column 3.
\end{flushleft}
\end{table}

\section{Concluding Remarks}

We have estimated how the mean density of dust in the universe varies with redshift using
a submillimetre method that has been designed to minimise the effect of dust temperature. We have used the Herschel-ATLAS
survey to show that the median temperature of dust in galaxies is $\rm \simeq 22\ K$ and does not vary with
redshift out to $z=1$. We have then used this temperature and an 850-$\mu$m survey of the COSMOS field
to estimate the mean density of dust in 10 redshift bins over the range $0 < z < 5.5$.
The main uncertainty
is whether the dust temperature changes at $z>1$, although we 
have argued that both simulations and observations suggest that the mean mass-weighted
dust temperature does not vary much with redshift.
It will be possible to do a little better in the future
using surveys selected at $\rm >1 mm$, where the relationship between continuum emission and dust temperature is even
weaker, although it will never be possible to remove the bias towards warmer temperatures completely.

We find that the mean density of dust in the universe increased by a factor of $\simeq$10 from $z=5$ to $z=2$. It then declined
slightly to $z=1$, and then declined steeply to the present day. The relationship we find is quite similar to the star-formation
history of the universe (Fig. \ref{fig:dust density with sfr}), although the increase in mean dust density from $z=5$ to $z=2$ is steeper than the
increase in the star-formation rate over the same redshift range. 
We have also used the submillimetre measurements to estimate the mean density of
hydrogen over the same redshift range. We have compared our estimates with estimates in the literature of the
mean density of atomic hydrogen and of CO-traced molecular hydrogen. The relationship for atomic hydrogen has a much larger amplitude
and a very different redshift dependence than the relationship for our estimates of the gas density.
However, the relationship with redshift of our estimates of the gas density (the dust-traced gas) is very similar
to the redshift relationship estimated from CO (the CO-traced gas), which implies 
that most of the dust
in the universe is found in the molecular rather than the atomic phase. 

We find that the depletion time for the dust-traced gas in the universe as a whole declines with redshift in the same way
that is seen in the galaxy population \citep{tacconi2020}.

\section*{Acknowledgements}

This analysis was started while SAE was at the conference {\it New Views on Feedback and the Baryon Cycle
in Galaxies} in Healesville Australia in Summer 2023 and continued during a visit to Swinburne University. He thanks the
organisers of the conference and the people that made the visit possible. He thanks the Science
and Technology Facilities Councile (consolidated grant ST/K000926/1) and the Taith fund, Wales' international exchange
programme, for the funds for this visit. BW thanks the Science and Technology Facilities Council for a PhD studentship.
We thank the referee Matthieu Béthermin for useful comments.

\section*{Data Availability}

The data from the {\it Herschel} ATLAS, which is used in Section 2.1 is publicly available (h-atlas.org). The data
for the COSMOS survey was taken from two papers \citep{davidzon2017,Millard_2020}.



\bibliographystyle{mnras}
\bibliography{mnras} 

\begin{thebibliography}{}
\makeatletter
\relax
\def\mn@urlcharsother{\let\do\@makeother \do\$\do\&\do\#\do\^\do\_\do\%\do\~}
\def\mn@doi{\begingroup\mn@urlcharsother \@ifnextchar [ {\mn@doi@} {\mn@doi@[]}}
\def\mn@doi@[#1]#2{\def\@tempa{#1}\ifx\@tempa\@empty \href {http://dx.doi.org/#2} {doi:#2}\else \href {http://dx.doi.org/#2} {#1}\fi \endgroup}
\def\mn@eprint#1#2{\mn@eprint@#1:#2::\@nil}
\def\mn@eprint@arXiv#1{\href {http://arxiv.org/abs/#1} {{\tt arXiv:#1}}}
\def\mn@eprint@dblp#1{\href {http://dblp.uni-trier.de/rec/bibtex/#1.xml} {dblp:#1}}
\def\mn@eprint@#1:#2:#3:#4\@nil{\def\@tempa {#1}\def\@tempb {#2}\def\@tempc {#3}\ifx \@tempc \@empty \let \@tempc \@tempb \let \@tempb \@tempa \fi \ifx \@tempb \@empty \def\@tempb {arXiv}\fi \@ifundefined {mn@eprint@\@tempb}{\@tempb:\@tempc}{\expandafter \expandafter \csname mn@eprint@\@tempb\endcsname \expandafter{\@tempc}}}

\bibitem[\protect\citeauthoryear{{Barger}, {Cowie}, {Blair}  \& {Jones}}{{Barger} et~al.}{2022}]{Barger_2022}
{Barger} A.~J.,  {Cowie} L.~L.,  {Blair} A.~H.,   {Jones} L.~H.,  2022, \mn@doi [\apj] {10.3847/1538-4357/ac67e7}, \href {https://ui.adsabs.harvard.edu/abs/2022ApJ...934...56B} {934, 56}

\bibitem[\protect\citeauthoryear{{Bendo} et~al.,}{{Bendo} et~al.}{2023}]{Bendo_2023}
{Bendo} G.~J.,  et~al., 2023, \mn@doi [\mnras] {10.1093/mnras/stac3771}, \href {https://ui.adsabs.harvard.edu/abs/2023MNRAS.522.2995B} {522, 2995}

\bibitem[\protect\citeauthoryear{{B{\'e}thermin} et~al.,}{{B{\'e}thermin} et~al.}{2015}]{Bethermin_2015}
{B{\'e}thermin} M.,  et~al., 2015, \mn@doi [\aap] {10.1051/0004-6361/201425031}, \href {https://ui.adsabs.harvard.edu/abs/2015A&A...573A.113B} {573, A113}

\bibitem[\protect\citeauthoryear{{Bolatto}, {Wolfire}  \& {Leroy}}{{Bolatto} et~al.}{2013}]{Bolatto2013}
{Bolatto} A.~D.,  {Wolfire} M.,   {Leroy} A.~K.,  2013, \mn@doi [\araa] {10.1146/annurev-astro-082812-140944}, \href {https://ui.adsabs.harvard.edu/abs/2013ARA&A..51..207B} {51, 207}

\bibitem[\protect\citeauthoryear{{Bourne} et~al.,}{{Bourne} et~al.}{2016}]{bourne2016}
{Bourne} N.,  et~al., 2016, \mn@doi [\mnras] {10.1093/mnras/stw1654}, \href {https://ui.adsabs.harvard.edu/abs/2016MNRAS.462.1714B} {462, 1714}

\bibitem[\protect\citeauthoryear{{Cortese} et~al.,}{{Cortese} et~al.}{2014}]{Cortese2014}
{Cortese} L.,  et~al., 2014, \mn@doi [\mnras] {10.1093/mnras/stu175}, \href {https://ui.adsabs.harvard.edu/abs/2014MNRAS.440..942C} {440, 942}

\bibitem[\protect\citeauthoryear{{Davidzon} et~al.,}{{Davidzon} et~al.}{2017}]{davidzon2017}
{Davidzon} I.,  et~al., 2017, \mn@doi [\aap] {10.1051/0004-6361/201730419}, \href {https://ui.adsabs.harvard.edu/abs/2017A&A...605A..70D} {605, A70}

\bibitem[\protect\citeauthoryear{{Decarli} et~al.,}{{Decarli} et~al.}{2016}]{decarli2016}
{Decarli} R.,  et~al., 2016, \mn@doi [\apj] {10.3847/1538-4357/833/1/69}, \href {https://ui.adsabs.harvard.edu/abs/2016ApJ...833...69D} {833, 69}

\bibitem[\protect\citeauthoryear{{Decarli} et~al.,}{{Decarli} et~al.}{2019}]{decarli2019}
{Decarli} R.,  et~al., 2019, \mn@doi [\apj] {10.3847/1538-4357/ab30fe}, \href {https://ui.adsabs.harvard.edu/abs/2019ApJ...882..138D} {882, 138}

\bibitem[\protect\citeauthoryear{{Drew} \& {Casey}}{{Drew} \& {Casey}}{2022}]{Drew_2022}
{Drew} P.~M.,  {Casey} C.~M.,  2022, \mn@doi [\apj] {10.3847/1538-4357/ac6270}, \href {https://ui.adsabs.harvard.edu/abs/2022ApJ...930..142D} {930, 142}

\bibitem[\protect\citeauthoryear{{Driver} et~al.,}{{Driver} et~al.}{2018}]{driver2018}
{Driver} S.~P.,  et~al., 2018, \mn@doi [\mnras] {10.1093/mnras/stx2728}, \href {https://ui.adsabs.harvard.edu/abs/2018MNRAS.475.2891D} {475, 2891}

\bibitem[\protect\citeauthoryear{{Dudzevi{\v{c}}i{\={u}}t{\.{e}}} et~al.,}{{Dudzevi{\v{c}}i{\={u}}t{\.{e}}} et~al.}{2020}]{Dudzeviciute_2020}
{Dudzevi{\v{c}}i{\={u}}t{\.{e}}} U.,  et~al., 2020, \mn@doi [\mnras] {10.1093/mnras/staa769}, \href {https://ui.adsabs.harvard.edu/abs/2020MNRAS.494.3828D} {494, 3828}

\bibitem[\protect\citeauthoryear{{Dunne}, {Eales}, {Edmunds}, {Ivison}, {Alexander}  \& {Clements}}{{Dunne} et~al.}{2000}]{Dunne_2000}
{Dunne} L.,  {Eales} S.,  {Edmunds} M.,  {Ivison} R.,  {Alexander} P.,   {Clements} D.~L.,  2000, \mn@doi [\mnras] {10.1046/j.1365-8711.2000.03386.x}, \href {https://ui.adsabs.harvard.edu/abs/2000MNRAS.315..115D} {315, 115}

\bibitem[\protect\citeauthoryear{{Dunne} et~al.,}{{Dunne} et~al.}{2011}]{dunne2011}
{Dunne} L.,  et~al., 2011, \mn@doi [\mnras] {10.1111/j.1365-2966.2011.19363.x}, \href {https://ui.adsabs.harvard.edu/abs/2011MNRAS.417.1510D} {417, 1510}

\bibitem[\protect\citeauthoryear{{Dunne}, {Maddox}, {Papadopoulos}, {Ivison}  \& {Gomez}}{{Dunne} et~al.}{2022}]{Dunne_2022}
{Dunne} L.,  {Maddox} S.~J.,  {Papadopoulos} P.~P.,  {Ivison} R.~J.,   {Gomez} H.~L.,  2022, \mn@doi [\mnras] {10.1093/mnras/stac2098}, \href {https://ui.adsabs.harvard.edu/abs/2022MNRAS.517..962D} {517, 962}

\bibitem[\protect\citeauthoryear{{Dye} et~al.,}{{Dye} et~al.}{2010}]{dye2010}
{Dye} S.,  et~al., 2010, \mn@doi [\aap] {10.1051/0004-6361/201014614}, \href {https://ui.adsabs.harvard.edu/abs/2010A&A...518L..10D} {518, L10}

\bibitem[\protect\citeauthoryear{{Eales}, {Wynn-Williams}  \& {Duncan}}{{Eales} et~al.}{1989}]{eales1989}
{Eales} S.~A.,  {Wynn-Williams} C.~G.,   {Duncan} W.~D.,  1989, \mn@doi [\apj] {10.1086/167341}, \href {https://ui.adsabs.harvard.edu/abs/1989ApJ...339..859E} {339, 859}

\bibitem[\protect\citeauthoryear{{Eales} et~al.,}{{Eales} et~al.}{2010}]{Eales_2010}
{Eales} S.,  et~al., 2010, \mn@doi [\pasp] {10.1086/653086}, \href {https://ui.adsabs.harvard.edu/abs/2010PASP..122..499E} {122, 499}

\bibitem[\protect\citeauthoryear{{Eales} et~al.,}{{Eales} et~al.}{2012}]{Eales_2012}
{Eales} S.,  et~al., 2012, \mn@doi [\apj] {10.1088/0004-637X/761/2/168}, \href {https://ui.adsabs.harvard.edu/abs/2012ApJ...761..168E} {761, 168}

\bibitem[\protect\citeauthoryear{{Eales} et~al.,}{{Eales} et~al.}{2018}]{Eales_2018}
{Eales} S.,  et~al., 2018, \mn@doi [\mnras] {10.1093/mnras/stx2548}, \href {https://ui.adsabs.harvard.edu/abs/2018MNRAS.473.3507E} {473, 3507}

\bibitem[\protect\citeauthoryear{{Furlanetto} et~al.,}{{Furlanetto} et~al.}{2018}]{furlanetto2018}
{Furlanetto} C.,  et~al., 2018, \mn@doi [\mnras] {10.1093/mnras/sty151}, \href {https://ui.adsabs.harvard.edu/abs/2018MNRAS.476..961F} {476, 961}

\bibitem[\protect\citeauthoryear{{Garratt} et~al.,}{{Garratt} et~al.}{2021}]{Garratt2021}
{Garratt} T.~K.,  et~al., 2021, \mn@doi [\apj] {10.3847/1538-4357/abec81}, \href {https://ui.adsabs.harvard.edu/abs/2021ApJ...912...62G} {912, 62}

\bibitem[\protect\citeauthoryear{{Ismail} et~al.,}{{Ismail} et~al.}{2023}]{Ismail_2023}
{Ismail} D.,  et~al., 2023, \mn@doi [arXiv e-prints] {10.48550/arXiv.2307.15747}, \href {https://ui.adsabs.harvard.edu/abs/2023arXiv230715747I} {p. arXiv:2307.15747}

\bibitem[\protect\citeauthoryear{{James}, {Dunne}, {Eales}  \& {Edmunds}}{{James} et~al.}{2002}]{James_2003}
{James} A.,  {Dunne} L.,  {Eales} S.,   {Edmunds} M.~G.,  2002, \mn@doi [\mnras] {10.1046/j.1365-8711.2002.05660.x}, \href {https://ui.adsabs.harvard.edu/abs/2002MNRAS.335..753J} {335, 753}

\bibitem[\protect\citeauthoryear{{Liang} et~al.,}{{Liang} et~al.}{2019}]{Liang_2019}
{Liang} L.,  et~al., 2019, \mn@doi [\mnras] {10.1093/mnras/stz2134}, \href {https://ui.adsabs.harvard.edu/abs/2019MNRAS.489.1397L} {489, 1397}

\bibitem[\protect\citeauthoryear{{Lim} et~al.,}{{Lim} et~al.}{2020}]{Lim_2020}
{Lim} C.-F.,  et~al., 2020, \mn@doi [\apj] {10.3847/1538-4357/ab607f}, \href {https://ui.adsabs.harvard.edu/abs/2020ApJ...889...80L} {889, 80}

\bibitem[\protect\citeauthoryear{{Madau} \& {Dickinson}}{{Madau} \& {Dickinson}}{2014}]{madau2006}
{Madau} P.,  {Dickinson} M.,  2014, \mn@doi [\araa] {10.1146/annurev-astro-081811-125615}, \href {https://ui.adsabs.harvard.edu/abs/2014ARA&A..52..415M} {52, 415}

\bibitem[\protect\citeauthoryear{{Maddox} et~al.,}{{Maddox} et~al.}{2018}]{maddox2018}
{Maddox} S.~J.,  et~al., 2018, \mn@doi [\apjs] {10.3847/1538-4365/aab8fc}, \href {https://ui.adsabs.harvard.edu/abs/2018ApJS..236...30M} {236, 30}

\bibitem[\protect\citeauthoryear{{Magnelli} et~al.,}{{Magnelli} et~al.}{2014}]{Magnelli_2014}
{Magnelli} B.,  et~al., 2014, \mn@doi [\aap] {10.1051/0004-6361/201322217}, \href {https://ui.adsabs.harvard.edu/abs/2014A&A...561A..86M} {561, A86}

\bibitem[\protect\citeauthoryear{{M{\'e}nard} \& {Fukugita}}{{M{\'e}nard} \& {Fukugita}}{2012}]{menard2012}
{M{\'e}nard} B.,  {Fukugita} M.,  2012, \mn@doi [\apj] {10.1088/0004-637X/754/2/116}, \href {https://ui.adsabs.harvard.edu/abs/2012ApJ...754..116M} {754, 116}

\bibitem[\protect\citeauthoryear{{M{\'e}nard}, {Scranton}, {Fukugita}  \& {Richards}}{{M{\'e}nard} et~al.}{2010}]{menard2010}
{M{\'e}nard} B.,  {Scranton} R.,  {Fukugita} M.,   {Richards} G.,  2010, \mn@doi [\mnras] {10.1111/j.1365-2966.2010.16486.x}, \href {https://ui.adsabs.harvard.edu/abs/2010MNRAS.405.1025M} {405, 1025}

\bibitem[\protect\citeauthoryear{{Millard} et~al.,}{{Millard} et~al.}{2020}]{Millard_2020}
{Millard} J.~S.,  et~al., 2020, \mn@doi [\mnras] {10.1093/mnras/staa609}, \href {https://ui.adsabs.harvard.edu/abs/2020MNRAS.494..293M} {494, 293}

\bibitem[\protect\citeauthoryear{{Millard}, {Diemer}, {Eales}, {Gomez}, {Beeston}  \& {Smith}}{{Millard} et~al.}{2021}]{Millard2021}
{Millard} J.~S.,  {Diemer} B.,  {Eales} S.~A.,  {Gomez} H.~L.,  {Beeston} R.,   {Smith} M. W.~L.,  2021, \mn@doi [\mnras] {10.1093/mnras/staa3207}, \href {https://ui.adsabs.harvard.edu/abs/2021MNRAS.500..871M} {500, 871}

\bibitem[\protect\citeauthoryear{{P{\'e}roux} \& {Howk}}{{P{\'e}roux} \& {Howk}}{2020}]{peroux2020}
{P{\'e}roux} C.,  {Howk} J.~C.,  2020, \mn@doi [\araa] {10.1146/annurev-astro-021820-120014}, \href {https://ui.adsabs.harvard.edu/abs/2020ARA&A..58..363P} {58, 363}

\bibitem[\protect\citeauthoryear{{Planck Collaboration} et~al.,}{{Planck Collaboration} et~al.}{2015}]{Planck_2015}
{Planck Collaboration} et~al., 2015, \mn@doi [\aap] {10.1051/0004-6361/201424088}, \href {https://ui.adsabs.harvard.edu/abs/2015A&A...576A.107P} {576, A107}

\bibitem[\protect\citeauthoryear{{Pozzi}, {Calura}, {Zamorani}, {Delvecchio}, {Gruppioni}  \& {Santini}}{{Pozzi} et~al.}{2020}]{pozzi2020}
{Pozzi} F.,  {Calura} F.,  {Zamorani} G.,  {Delvecchio} I.,  {Gruppioni} C.,   {Santini} P.,  2020, \mn@doi [\mnras] {10.1093/mnras/stz2724}, \href {https://ui.adsabs.harvard.edu/abs/2020MNRAS.491.5073P} {491, 5073}

\bibitem[\protect\citeauthoryear{{R{\'e}my-Ruyer} et~al.,}{{R{\'e}my-Ruyer} et~al.}{2014}]{remy-ruyer2014}
{R{\'e}my-Ruyer} A.,  et~al., 2014, \mn@doi [\aap] {10.1051/0004-6361/201322803}, \href {https://ui.adsabs.harvard.edu/abs/2014A&A...563A..31R} {563, A31}

\bibitem[\protect\citeauthoryear{{Riechers} et~al.,}{{Riechers} et~al.}{2019}]{riechers2019}
{Riechers} D.~A.,  et~al., 2019, \mn@doi [\apj] {10.3847/1538-4357/aafc27}, \href {https://ui.adsabs.harvard.edu/abs/2019ApJ...872....7R} {872, 7}

\bibitem[\protect\citeauthoryear{{Schreiber}, {Elbaz}, {Pannella}, {Ciesla}, {Wang}  \& {Franco}}{{Schreiber} et~al.}{2018}]{Schreiber_2018}
{Schreiber} C.,  {Elbaz} D.,  {Pannella} M.,  {Ciesla} L.,  {Wang} T.,   {Franco} M.,  2018, \mn@doi [\aap] {10.1051/0004-6361/201731506}, \href {https://ui.adsabs.harvard.edu/abs/2018A&A...609A..30S} {609, A30}

\bibitem[\protect\citeauthoryear{{Scoville} et~al.,}{{Scoville} et~al.}{2014}]{Scoville_2014}
{Scoville} N.,  et~al., 2014, \mn@doi [\apj] {10.1088/0004-637X/783/2/84}, \href {https://ui.adsabs.harvard.edu/abs/2014ApJ...783...84S} {783, 84}

\bibitem[\protect\citeauthoryear{{Simpson} et~al.,}{{Simpson} et~al.}{2017}]{Simpson_2017}
{Simpson} J.~M.,  et~al., 2017, \mn@doi [\apj] {10.3847/1538-4357/aa65d0}, \href {https://ui.adsabs.harvard.edu/abs/2017ApJ...839...58S} {839, 58}

\bibitem[\protect\citeauthoryear{{Smith} et~al.,}{{Smith} et~al.}{2010}]{Smith2010}
{Smith} M.~W.~L.,  et~al., 2010, \mn@doi [\aap] {10.1051/0004-6361/201014584}, \href {https://ui.adsabs.harvard.edu/abs/2010A&A...518L..51S} {518, L51}

\bibitem[\protect\citeauthoryear{{Smith} et~al.,}{{Smith} et~al.}{2012}]{Smith2012}
{Smith} M.~W.~L.,  et~al., 2012, \mn@doi [\apj] {10.1088/0004-637X/756/1/40}, \href {https://ui.adsabs.harvard.edu/abs/2012ApJ...756...40S} {756, 40}

\bibitem[\protect\citeauthoryear{{Smith} et~al.,}{{Smith} et~al.}{2017}]{smith2017}
{Smith} M. W.~L.,  et~al., 2017, \mn@doi [\apjs] {10.3847/1538-4365/aa9b35}, \href {https://ui.adsabs.harvard.edu/abs/2017ApJS..233...26S} {233, 26}

\bibitem[\protect\citeauthoryear{{Tacconi}, {Genzel}  \& {Sternberg}}{{Tacconi} et~al.}{2020}]{tacconi2020}
{Tacconi} L.~J.,  {Genzel} R.,   {Sternberg} A.,  2020, \mn@doi [\araa] {10.1146/annurev-astro-082812-141034}, \href {https://ui.adsabs.harvard.edu/abs/2020ARA&A..58..157T} {58, 157}

\bibitem[\protect\citeauthoryear{{Valiante} et~al.,}{{Valiante} et~al.}{2016}]{valiante2016}
{Valiante} E.,  et~al., 2016, \mn@doi [\mnras] {10.1093/mnras/stw1806}, \href {https://ui.adsabs.harvard.edu/abs/2016MNRAS.462.3146V} {462, 3146}

\bibitem[\protect\citeauthoryear{{Vlahakis}, {Dunne}  \& {Eales}}{{Vlahakis} et~al.}{2005}]{Vlahakis2005}
{Vlahakis} C.,  {Dunne} L.,   {Eales} S.,  2005, \mn@doi [\mnras] {10.1111/j.1365-2966.2005.09666.x}, \href {https://ui.adsabs.harvard.edu/abs/2005MNRAS.364.1253V} {364, 1253}

\bibitem[\protect\citeauthoryear{{Walter} et~al.,}{{Walter} et~al.}{2014}]{walter2014}
{Walter} F.,  et~al., 2014, \mn@doi [\apj] {10.1088/0004-637X/782/2/79}, \href {https://ui.adsabs.harvard.edu/abs/2014ApJ...782...79W} {782, 79}

\bibitem[\protect\citeauthoryear{{Ward} et~al.,}{{Ward} et~al.}{2022}]{ward2022}
{Ward} B.~A.,  et~al., 2022, \mn@doi [\mnras] {10.1093/mnras/stab3300}, \href {https://ui.adsabs.harvard.edu/abs/2022MNRAS.510.2261W} {510, 2261}

\bibitem[\protect\citeauthoryear{{Witstok}, {Jones}, {Maiolino}, {Smit}  \& {Schneider}}{{Witstok} et~al.}{2023}]{Witstok_2023}
{Witstok} J.,  {Jones} G.~C.,  {Maiolino} R.,  {Smit} R.,   {Schneider} R.,  2023, \mn@doi [\mnras] {10.1093/mnras/stad1470}, \href {https://ui.adsabs.harvard.edu/abs/2023MNRAS.523.3119W} {523, 3119}

\bibitem[\protect\citeauthoryear{{Zavala} et~al.,}{{Zavala} et~al.}{2018}]{Zavala_2018}
{Zavala} J.~A.,  et~al., 2018, \mn@doi [\mnras] {10.1093/mnras/sty217}, \href {https://ui.adsabs.harvard.edu/abs/2018MNRAS.475.5585Z} {475, 5585}

\bibitem[\protect\citeauthoryear{{da Cunha}, {Charlot}  \& {Elbaz}}{{da Cunha} et~al.}{2008}]{dacunha2008}
{da Cunha} E.,  {Charlot} S.,   {Elbaz} D.,  2008, \mn@doi [\mnras] {10.1111/j.1365-2966.2008.13535.x}, \href {https://ui.adsabs.harvard.edu/abs/2008MNRAS.388.1595D} {388, 1595}

\bibitem[\protect\citeauthoryear{{da Cunha} et~al.,}{{da Cunha} et~al.}{2013}]{daCunha_2013}
{da Cunha} E.,  et~al., 2013, \mn@doi [\apj] {10.1088/0004-637X/766/1/13}, \href {https://ui.adsabs.harvard.edu/abs/2013ApJ...766...13D} {766, 13}

\makeatother
\end{thebibliography}




\appendix


\bsp	
\label{lastpage}
\end{document}